\documentclass[article]{JHEP}
\usepackage{amsmath,epsfig}
\usepackage{amssymb,amsfonts}
\usepackage{epsfig}
\relax
\renewcommand{\theequation}{\arabic{section}.\arabic{equation}}
\newcommand{\nn}{\nonumber}

\newcommand{\de}{\partial}

\def\simleq{\; \raise0.3ex\hbox{$<$\kern-0.75em\raise-1.1ex\hbox{$\sim$}}\; }
\def\simgeq{\; \raise0.3ex\hbox{$>$\kern-0.75em\raise-1.1ex\hbox{$\sim$}}\; }

\def\slash{\raise.15ex\hbox{$/$}\kern-.57em}
\def\slashA{\raise.15ex\hbox{$/$}\kern-.70em A}

\def\l{\lambda}

\def\r{\rho}

\newcommand{\beq}{\begin{equation}}
\newcommand{\eeq}{\end{equation}}
\newcommand{\bea}{\begin{eqnarray}}
\newcommand{\eea}{\end{eqnarray}}

\title{Metastable Vacua in Brane Worlds}
\author{
E. Dudas$^{1,2}$, J. Mourad$^{3}$, F. Nitti$^{1}$\\
$^1$CPHT, Ecole Polytechnique, CNRS,
 91128, Palaiseau, France\\
 ( UMR du CNRS 7644)
~\\
$^2$ LPT-Orsay, Bat. 210, Univ. Paris-Sud, 91405 Orsay Cedex, France\\
(UMR du CNRS 8627)\\
$^3$ APC, Univ. Paris VII, Bat. Condorcet, 75205 Paris Cedex 13, France \\
~\\
}
\preprint{arXiv:0706.1269 \\ CPHT-RR044.0607 \\LPT-Orsay 07-37 }

\abstract{We analyze vacuum decay  in brane world setups, where a
free scalar field in five dimensions has a localized
potential admitting metastable vacua. We  study in particular the
bounce solution and its properties in flat and warped spaces. In the
latter case, placing into a deeply warped region the term in the potential 
that lifts  the vacuum degeneracy,  
can increase indefinitely the lifetime of the false
vacuum. We discuss the application to metastable vacua in
supersymmetric brane-world constructions.}

\begin{document}

\maketitle 

\section{Introduction and conclusions}

The fate of metastable vacua in field theory \cite{coleman,clucia} is of
great interest in cosmology and particle physics. The dynamics of
their quantum decay toward the true vacuum rely on the knowledge of
the classical Euclidean ``bounce'' solution. The study of finite
energy soliton solutions in Minkowski space and of finite action
solutions in Euclidean space is therefore crucial for the study of
metastable vacua. The purpose of this paper is to study the
generalization of this problem in the case of vacua generated by a
scalar field living in a higher-dimensional spacetime, with a scalar
potential localized in four dimensions. The field is therefore free
in the bulk, with the scalar potential generating non-trivial
boundary conditions. One of our main motivation for studying the
case of a boundary potential is the generalization to supersymmetric
theories with metastable vacua \cite{iss,issstrings}. In this case,
constraints coming from higher-dimensional supersymmetry are such
that it is much easier to construct models with localized (as opposed to
bulk)
superpotentials.

The search for classical solutions in this case
turns out to be very interesting and rich. A first question is the
dependence of the classical solution on the geometry of the internal
space, flat or warped, and on its size. In the case of the bounce,
we would like to understand the dependence of the width of the wall
in the thin wall approximation on the extra dimension.  Moreover,
since all nontrivial dynamics is encoded in boundary conditions, it
suggests that the problem could be tractable to some extent even in
the case where the internal space is warped. In this last case,
there are several interesting questions arising. First of all, even
if we are in a regime in which the 4d effective theory is valid,
i.e. there is a mode much lighter than the KK masses, it is possible
that the barrier separating the false from the true minimum is much
higher than the mass of the lowest-lying KK states. In this case,
despite the validity of the 4d effective action describing the
lightest mode, there is no a priori reason why the classical solution
should be the standard 4d one. Secondly, it is reasonable to expect
that by placing the term lifting the degeneracy between the true and
the false vacuum into a deeply warped region, it will be redshifted
to small values, thus increasing indefinitely the lifetime of the
false vacuum. If this were indeed possible, there would be no
practical difference between living in the true vacuum or the false
vacuum ! In this paper, we will be able to answer some of these
questions, whereas other questions will be addressed only partially
and will need future work for a complete understanding.

The plan of the paper is as follows. In Section \ref{flat} we
discuss vacuum decay in  a toy-model consisting of a single scalar
field  in a spacetime with one flat, compact extra-dimension. We
show that when there is a light (compared to the KK scale) mode, the
effective theory is precisely the one which admits a standard kink
solution. We then work out the 5d analog of the classical 4d field
equation. This can be written as a 4d differential equation, which
allows a systematic calculation of the corrections to the 4d kink
solution. The equation involves a differential operator containing
higher derivative terms. We work out the size of the kink in the
large radius limit, show that the kink become broader with
increasing the radius and study its behavior near the origin. We use
this to write the equation defining the euclidian bounce and check
the validity of the thin wall approximation, finding that it gets
worse in the 5d limit. 

In Section \ref{warped} we perform the same
analysis including the effect of the warping of the extra-dimension.
We find that the 4d limit and the thin-wall approximation become
very accurate because of the warping. However Coleman-de Luccia
gravitational effects can become important and even completely lock
the decay of the false vacuum.  

In Section \ref{susy} we extend
these considerations to the supersymmetric case. Motivated by the
D3/D7 brane realizations of the ISS model, we discuss the $AdS_5$
version of the ISS model \cite{iss}, with ISS gauge group and quarks
living on the UV boundary and the mesons living in the 5d bulk. The
meson-quark coupling is then localized on the UV brane, whereas the
mesonic linear term in superpotential is put on the IR boundary. We
find that, due to the warping, the (mass)$^2$ parameter is naturally
redshifted to small values, whereas metastable supersymmetry
breaking becomes a non-local (in the extra dimension) effect. This
has again the net effect of increasing correspondingly the lifetime
of the metastable vacuum. We also analyze briefly the case where the
whole superpotential is localized on the UV brane and a light mode
is achieved by adding a bulk mass term for the mesons
hypermultiplet. In this case mass scales are redshifted again due to
a different effect, the value of the mesonic wave function on the UV
brane. Both examples have a natural 4d holographic interpretation
via the
 AdS/CFT correspondence.

Some technical details of the computations are left to three Appendices.


\section{Flat 4+1 Dimensions} \label{flat}

In this section we consider a massless scalar field in a 4+1
dimensional flat spacetime\footnote{We use signature $(-++++)$.
Throughout the paper we consistently neglect the backreaction of the
scalar field on the geometry \cite{GW,freedman}} in which the 5th
direction (labeled by the coordinate $y$) extends between two rigid
branes at $y=0$ and $y=\pi R$. The bulk action is that of a free
massless field, all nontrivial potential terms appearing on the
boundaries\footnote{The case of the bulk potential and the kink solution
in the extra coordinate did lead historically to the first brane world
proposals \cite{valery}. For bounce solutions for brane localized fields, see
e.g. \cite{db}.}
: \beq\label{action0} S = -{1\over 2} \int d^4 x dy\,\de_A
\Phi \de^A \Phi - \int d^4 x \left.V_0(\Phi)\right|_{y=0} +  \int
d^4 x \left.V_1(\Phi)\right|_{y=\pi R}. \eeq The field equations and
boundary conditions read: \bea
&& \de_y^2 \Phi + \de^\mu \de_\mu\Phi =0, \label{fe} \\
&& \left.\de_y \Phi\right|_{y=0} = {\de V_0 \over \de\Phi}, \label{bc0} \\
&& \left.\de_y \Phi\right|_{y=\pi R} = {\de V_1 \over \de\Phi}. \label{bc1}
\eea

\subsection{The Kink}

Consider the situation where the brane potentials are given by:
\beq\label{quartic}
V_0(\Phi) = {\l\over 4}\left(\Phi^2-v^2\right)^2, \qquad V_1(\Phi)=0
\eeq
From  eqs. (\ref{fe}-\ref{bc1}) we see immediately that there are two
``vacuum'' solutions $\Phi_{\pm} (x,y) = \pm v$. One can ask whether there
exist a solution interpolating between the two vacua, analogous to the
four-dimensional domain wall (kink) that one finds with the same
quartic potential (see Appendix \ref{appkink}).

Notice that, since $\Phi$ is canonically normalized in 5D, and has
mass dimension 3/2,  the parameters in (\ref{quartic}) have unusual
mass dimensions: \beq\label{dims} [\l] = M^{-2}, \qquad  [v] =
M^{3/2}. \eeq
\subsubsection{Effective 4D theory}
A kink-like solution is expected to exist at least in
a certain region of parameter space, where one can give a four-dimensional
effective description of the model. To see this, consider the linearized
fluctuations around one of the two vacua (say $\Phi_-$):
\beq
\Phi(x,y) = -v + \delta\Phi(x,y).
\eeq
 Decomposing the solution in eigenstates of the 4D D'Alambertian,
$\delta\Phi(x,y) = \phi(y)\chi(x)$,   $\Box_4 \chi(y) = m^2  \chi(y)$,
 the mass spectrum is obtained by  linearizing the boundary conditions
(\ref{bc0}-\ref{bc1}):
\beq
\left[\de_y\Phi = \mu_0^2 \Phi\right]_{y=0} , \qquad \left[\de_y\Phi = 0\right]_{y=\pi R}
\eeq
where $\mu_0^2 = 2\l v^2$. The mass eigenstates are the solutions of
the  equation:
\beq\label{eigen}
m \tan  m \pi R = \mu_0^2
\eeq
and the profile wave-function for a given mode of mass $m$ is:
\beq\label{profile}
\phi_m(y)  = \cos [m (y-\pi R)].
\eeq
We have a low-energy, 4D effective theory for the lowest-lying mode
 (of mass $m_0$)  if $m_0  R \ll 1$. This description is valid
for energies much smaller than  the mass of the next KK mode, which
is of order $1/R$. Under this conditions we can expand the tangent
in eq. (\ref{eigen}) and obtain:
\beq
m_0^2 \simeq {\mu_0^2 \over \pi R} = {2\l v^2 \over \pi R}
\eeq
and the condition for the existence of a 4D description reads, in
terms of the original parameters of the model:
\beq\label{4d}
  2 R  \l v^2 \ll \pi \ .
\eeq

Under these condition, inserting  $\Phi(x,y) = - v + \phi_0(y)\chi_0(x)$ in
the original action and integrating over $y$, we obtain the low-energy
4D effective action for the lowest-lying mode $\chi_0(x)$. After some
integration by parts and using the bulk field equation we obtain:
\beq\label{eff}
S_{eff} = \int d^4x \left[ -{1\over 2} \de_\mu \chi_0 \de^\mu \chi_0 - V_{eff}(\chi_0)\right]
\eeq
where the effective potential is:
\bea\
&& V_{eff} = {m_0^2 \over 2} \chi_0^2 - {g\over 3} \chi_0^3 + {h\over 4}\chi_0^4; \label{veff} \\
&& m_0^2\simeq {2\l v^2 \over \pi R}, \qquad g = {3 \l v \over (\pi R)^{3/2}}, \qquad h = {\l\over (\pi R)^2}.
\eea
The extra factors of $\pi R$ in the effective parameters come from the
normalized wave-function profile $\phi_0(y) =( 1/\sqrt{\pi R} )\cos m_0(y-\pi R)$
evaluated in $y=0$. It is easy to check that the potential (\ref{veff}) has
two zero-energy minima at $\chi_0=0, 2 v \sqrt{\pi R}$  and a maximum at
$\chi_0= v\sqrt{\pi R}$ with $V(v \sqrt{\pi R}) = \l v^4/4$. In terms of the original
field $\Phi = -v + \phi_0 \chi_0 $
these correspond exactly to the original two minima at $\Phi=\pm v$ and
maximum at $\Phi=0$.

Due to the standard double-well form of the effective
potential  the field equation derived from the effective action (\ref{eff}),
 \beq
\de_x^2 \chi(x) = {\de V_{eff}\over \de \chi}
\eeq
admits a
kink solutions interpolating between the two vacua  $\chi_0=0$ and
$\chi_0 = 2v \sqrt{R}$, which  according to eqs. (\ref{intro1}, \ref{intro2})
has the form:
\beq\label{kink}
\chi_{kink}(x) = v\sqrt{\pi R} \left(1  + \tanh \mu x\right),
\quad \mu^2  = {h\over 2} \left(v\sqrt{\pi R}\right)^2 =
{\mu_0^2\over 4\pi R} =  {\l v^2\over 2 \pi R}.
\eeq

The kink  energy density  is of  the order of the height of the potential barrier,
$\l v^4$. In order for the solution we found to be reliable, this energy density must be
below the KK scale, we thus have the additional requirement $ \l v^4
\ll 1/R^4$. This, together with  (\ref{4d}), sets the range of validity of the kink solution
we found.  A sufficient condition is :
\beq\label{4d2}
v \ll R^{-3/2}, \qquad \l \ll R^{2}.
\eeq

Although the energy density of the domain wall is $R$-independent,
the integrated total energy is not:
\beq
E =\int_{-\infty}^{+\infty}
dx \left({d\chi_{kink} \over dx}\right)^2 \sim v^2 \, \pi R \,\mu =
\sqrt{\l \pi R }\, v^3
\eeq
\subsubsection{The 5D equation} \label{5Deq}

The argument of the previous subsection suggest that a kink solution
to the model (\ref{action0}) should
exists, at least in the range of parameters satisfying (\ref{4d2}).
Now we want to look for similar solutions from a purely 5D perspective,
without having to rely on the 4D effective theory approach.

Let us return to eqs. (\ref{fe}-\ref{bc1}). We look for solutions
depending on $y$ and one of the Minkowski coordinates (say $x$). The
most general (real) solution to (\ref{fe}) can be written as:
\beq
\Phi(x,y) = g(x+iy) + (g(x+iy))^*.
\eeq
The boundary condition at $y=\pi R$ tells us that
\beq
Im \left[g'(x+i \pi R)\right]=0,
\eeq
where a prime denotes derivative w.r.t. the argument. This equation is
satisfied if
$F(x) = g(x+ i \pi R )$ is a real function\footnote{That is, $g(z)$ has an expansion
of the form
\beq
g(z) = \sum c_n (z-i\pi R)^n \nn
\eeq
with real coefficients $c_n$.}. This also implies that, for  $z$ complex, $(F(z))^*= F(z^*)$.

Next, consider the boundary conditions at $y=0$.
  Defining  $f(x) = \Phi(x,0)$, $h(x) = \de_y \Phi(x,0)$, eq. (\ref{bc0})
 reads:
\beq\label{bc02}
h(x)  = \l f(x) \left[\left(f(x)\right)^2 -v^2\right].
\eeq
 Formally, we can write:
\bea
 f(x)&&  = g(x) + g(x)^* = F(x-i\pi R) + F(x+i\pi R) \nn \\
&& = \Big(\exp[-i \pi R \de_x] +  \exp[i \pi R \de_x]\Big) F(x), \\
h(x) && = ig'(x) - i g'(x)^*  =  i  F'(x-i\pi R) - i F'(x+i\pi R) \nn\\
&&= i\Big(\exp[-i \pi R \de_x] -  \exp[i \pi R \de_x]\Big) \de_x F(x)\nn \\
&& = \tan (\pi R \de_x) \de_x f(x).\label{h}
\eea
Using the last line in eq. (\ref{h}) we arrive at a closed equation\footnote{As for the  standard kink,
this equation can be obtained from a the point-particle analog model,
with potential $-V(f)$
and   an exotic kinetic term:
\beq\label{effaction}
S = \int dx \left[V(f) - {1\over 2}  f\tan (\pi R \de_x )\de_x f\right],
\eeq
which is the same as the ``effective action''  whose variation gives (\ref{tan}).

}
for $f(x)$:
\\

\beq\label{tan}
 \tan (\pi R \de_x) \ \de_x f =  \l f \ \left[f^2 -v^2\right] \ .
\eeq
\\

Another, maybe less general way, in order to arrive at (\ref{tan})
is to start from the bulk solution \beq \Phi (x,y) = \int dp \ a_p \
e^{p x} \ \cos (p y + \alpha_p ) \ , \label {flat1} \eeq where $a_p$
($\alpha_p$) are arbitrary coefficients (phases). Boundary
conditions at $y = \pi R$ fixes $\alpha_p = - p \ \pi R$, whereas
boundary conditions in $y=0$ gives by a straightforward computation
(\ref{tan}), by using the replacement $p \rightarrow \de_x$. This
method will generalize in a straightforward  manner to the warped case
discussed in the next section.

 A solution to (\ref{tan}) gives $f(x) = \Phi(x,0)$, which
can then be extended into the bulk to a full solution: \bea
\label{bulksol} && \Phi(x,y) =Re [f(x+ i y)] + \tan (\pi R \de_x)
Im[f(x+iy)]
\nonumber \\
&& = \left[  \cos y \de_x + (\tan \pi R \de_x) (\sin y \de_x)
\right] f (x) \ . \eea

Eq. (\ref{tan}) has various interesting properties. It should be understood
as a series in  derivatives of increasing order. If we take $f(x)$ to be
a 4D mass eigenstate, $f(x) = e^{m x}$, and linearize the r.h.s, we get back
to the eigenvalue equation (\ref{eigen}). So the information about
the spectrum of the model is contained in (\ref{tan}).

Now suppose that we can keep the lowest order in the expansion of
the l.h.s. (for any given solution we can later check whether this
approximation is justified). We get a second order equation for $f$
which looks exactly as the one for the 4D kink:
\beq \label{kinkeq2}
\de_x^2 f \ = \ {\l\over \pi R} f \left[f^2 - v^2 \right]
\eeq
whose
solution is again given by eq. (\ref{intro2}): \beq \label{kink2}
f(x) \ = \ v \ \tanh \mu x,  \qquad \mu^2  = \l v^2/(2 \pi R). \eeq
Notice that  the characteristic scale $\mu$ is the same as in eq.
(\ref{kink}).

We can extract considerable information from eq.  (\ref{tan}) even when
the 4D limit does not hold. Consider a solution $f(x)$ that
approaches $\pm v$ as $x\to \pm \infty$. We can estimate the width
of the kink by expanding $f(x) = -v + \eta(x)$ and solving the
linear equation for $\eta$ in the asymptotic large $|x|$ region:
assuming $\eta(x) \sim e^{-|x|/l_w}$, where $l_w$ is a measure of
the wall width,   we find:
\beq\label{lw}
{1\over l_w} \ \tan \left({\pi
R\over l_w}\right) \ = \ 2 \l v^2 \ .
\eeq
In the 4D limit we get
the expected result, namely $l_w = 1/\mu$, with $\mu$ as in
(\ref{kink2}). In any case, the size of the wall cannot exceed $2R$,
and this value is approached in the opposite limit, when $R\l
v^2>>1$.

Another interesting length scale is the one corresponding to the
regime of the validity of the linear slope of the solution $f(x)
\sim (1 / l_0) x$, when the variation (derivative) of the field
$f$ is maximal\footnote{In the usual $d+1$ kink solution
(\ref{intro2}) both $l_{w}$ and $l_0$ are of order $1/\mu$.}. This
can be estimated by linearizing eq. (\ref{tan}) around $f=0$.
Setting  $f (x) = \eta \sin (x/l_0)$, with $\eta$ a constant, we find:
\beq \label{l0}
{1\over l_0} \ \tanh \left({\pi R\over l_0}\right) \ = \
\l v^2 \ .
 \eeq
In the 4d limit we get as expected $l_0 \sim 1/\mu$, whereas in the
5d regime  we get $l_0 \sim (1/\l v^2)$. In the 5d limit therefore,
the size of the kink becomes larger and larger, whereas large
variations of the field are confined into a fixed region.

We can put eq. (\ref{tan}) in integral form. Going in Fourier space,
and using the identity \beq \int dk {e^{ik x}\over k \tanh k}  \ = -
\  \log [\sinh |\pi x/2 |], \eeq and $f(0)=0$, we obtain:
\beq\label{int} \hat{f}(u) = {1\over 2\pi} \int  dt \, \log
\left[\sinh  \left|{u - t\over 2R\l v^2}\right|\right]
\hat{f}(t)(\hat{f}(t)^2 -1) \eeq where we have defined  $\hat{f}(u)
\equiv v^{-1} f(u/(\l v^2))$. Since the kernel in (\ref{int})
behaves as $|x-t|$ for large $t$, $f(t)$ must necessarily approach
one of the extrema $ f =0,\pm v$ as $t\to \pm \infty$.

It is also of interest to find the generalization of equ. (\ref{intro3})
which expresses the vanishing of the "energy" of the point particle analog
system. The determination of this quantity turns out to be non-trivial.
The details of the calculation are relegated to the Appendix where we show that
the kink verifies the following equation
\beq
{\cal E}={1\over 2}{\tan(\pi R\partial) \over \pi R \partial}
\left[(\partial f)^2-(\tan(\pi R\partial)\partial f)^2\right]-{1\over \pi
R}V(f)=0.\label{ene}
\eeq
The leading terms in the expansion in powers of $R$ are
\beq
 {1\over 2}(\partial f)^2-{1\over \pi R}V(f)+\dots=0.
\eeq They reproduce the 4D equation (\ref{intro3}). The first order
correction to the 4D equation of motion can also be simply obtained:
\beq {1\over 2}[(\partial f)^2+{1\over 3}(\pi
R)^2(2\partial^3f\partial f -(\partial^2f)^2)] -{1\over \pi
R}V(f)=0. \eeq If we expand $f$ in powers of $R$ and write
$f=f_0+(\pi R)^2 f_2+\dots$, then $f_2$ can be determined from the
first order equation: \beq f_0'f_2'-{1\over \pi R}V'(f_0)f_2+{1\over
6}(2\partial^3f_0\partial f_0 -(\partial^2f_0)^2)=0,\label{first}
\eeq where $f_0$ is the 4D solution $v\tanh\mu x$. Using the zeroth
order equation of motion $f_0''={1\over \pi R}V'(f_0)$ the solution
can be written as \beq f_2(x)={f_0'(x)\over 6}\int ^x_0du
{((\partial^2f_0)^2 -2\partial^3f_0\partial f_0)\over (\partial
f_0)^2}, \eeq where the integration constant was fixed by requiring
$f_2$ to be odd. Finally the integration can be  done to yield \beq
f_2(x)={1\over 6}f_0'(x)\left[-{2\l\over {\pi R}}v^2x +
4\sqrt{2\l\over {\pi R}}f_0(x)\right]. \eeq Explicitly, the solution
to the first nontrivial order  reads \beq f(x)=v\tanh(\mu x)\left[1
- {(2\pi R\mu)^2\over 6\cosh^2{(\mu x)}}\left({\mu x\over\tanh(\mu
x)} - 2\right)\right]. \eeq
 This shows that the zeroth order approximation is valid as long as $(2\pi
 R\mu)^2$
 is much smaller than one.

The identification of the kink as a solution of the equation ${\cal E}=0$,
has an important consequence: such solutions can never cross
the lines $f =\pm v$. This is analog to the usual two-derivative kink:
there, eq. (\ref{intro3}) implies that $f=\pm v$ are the fixed points
fot the first order flow of the quantity  $\Phi$,
and as such they cannot be crossed in finite ``time.''
The corresponding statement in the case of eq. (\ref{ene}) is
prooven in Appendix (\ref{fixed}). As an important consequence of this
fact, the (true) energy of any solution interpolating between $+v $ and $-v$    is  always positive, as we will see in the next subsection.

Another useful form of ${\cal E}$ is obtained by using the equations
of motion (\ref{tan}) in (\ref{ene}) to put it in the form
\beq
{\cal E}={1\over 2}{\tan(\pi R\partial) \over \pi R \partial}
\left[(\partial f)^2-(V'(f))^2\right]-{1\over \pi
R}V(f)=0.\label{ener}
\eeq

\subsection{The Bounce}

Next, we add a linear potential
on the brane at $y=\pi R$, of the form
\beq \label{linear}
V_1 = b (\Phi-v),   \qquad [b] = M^{5/2}.
\eeq
In this case, eq. (\ref{tan}) becomes:
\beq\label{tan1}
 \tan (\pi R \de_x) \ \de_x f - b \ = \  \l \ f \ \left[f^2 -v^2\right] \ .
\eeq

The addition of (\ref{linear}) breaks the $\Phi\to -\Phi$ symmetry and lifts the degeneracy
between the two vacua, making one of them metastable. We will follow
Coleman \cite{coleman} and estimate the
decay rate of the metastable vacuum. We  look for a  ``bounce'' solution
$\Phi^B(y,t_E, \vec{x})$, i.e. a  solution of the
Euclidean field equations that interpolates between
the true vacuum at  small $\rho \equiv \sqrt{t_E^2 + |\vec{x}|^2}$,
and the false vacuum at large $\r$. The bounce must have finite action
relative to the false vacuum.
Then the tunneling amplitude is given by:
\beq\label{rate}
\Gamma = \exp\left\{- S_E[\Phi^B] + S_E[\Phi^F]\right\}.
\eeq
Here  $ S_E[\Phi]$ is the euclidean version of the  action (\ref{action0}) evaluated
on the field configuration $\Phi^B$ (the bounce) and $\Phi^F$ (the false
vacuum).

We look for a bounce solution with  $O(4)$-symmetry, i.e. depending only on $y$ and on the Euclidean radial coordinate $\rho$.
The euclidean field equation in these coordinates reads:
\bea
&& \de_y^2 \Phi + \de_\rho^2 \Phi + {3\over \r} \de_\r \Phi =0 \label{feb}\\
&& \left.\de_y \Phi\right|_{y=0} = \l \Phi (\Phi^2-v^2), \label{bc0b} \\
&& \left.\de_y \Phi\right|_{y=\pi R} = b . \label{bc1b}
\eea
and we look for a solution that approaches the true vacuum $\Phi^T$ at
$\rho\simeq 0$ and the false vacuum $\Phi^F$ at $\rho \simeq \infty$.
Following Coleman, we consider the symmetry  breaking term as a perturbation:
 we approximate both the true and the false vacuum to be the
same as the unperturbed ones ($\Phi(x,y) = \pm v$), and moreover
we set $b=0$  when solving the field equation.  As a further approximation,
we assume we are in the ``thin wall'' limit, in which we can neglect
the last term in eq. (\ref{feb}). This is justified when the
transition between the true and false vacuum takes place in a
in a small region of width $l_b$ around a radius  $\rho_0 \gg l_b$. Under
these assumptions, the problem reduces to the one of the  previous
section, i.e. finding a domain wall  solution centered around $\rho_0$:
\beq\label{thinwall}
\Phi^B(\r,y) = \left\{ \begin{array}{ll} -v & 0<\r\ll \r_0, \\
                                        \Phi_{kink}(\r-\r_0,y)\quad  & \r \simeq \r_0 \\
                                         v & \r\gg \r_0, \end{array}\right.
\eeq

Requiring that the bounce has minimal action provides a variational problem
for the parameter $\r_0$.  The bounce action is:
\beq\label{sbounce0}
{S_b\over 2\pi^2} = \int_0^{\pi R} dy \int d\r \r^3 {1\over 2} \left[(\de_\r \Phi^B)^2  + (\de_y \Phi^B)^2\right] -   \int d\r \r^3 \left(V_0(\Phi^B(\r,0))
- V_1(\Phi^B(\r,\pi R))\right).
\eeq
Integrating by parts the bulk piece and  using the field equations
the above expression reduces to  boundary terms. Approximating
the solution as in  (\ref{thinwall}), we obtain:
\beq \label{sbounce}
{S_b\over 2\pi^2} \approx -2 b v {\rho_0^4\over 4} + \rho_0^3 S_{wall} \ .
\eeq
Here, $S_{wall}$ is the  energy stored in the wall.
In our approximation can be thought as concentrated in a small region around $\rho_0$, and can be approximated by the total energy of the kink, i.e. the solution of eq.  (\ref{tan}) :
\beq\label{swall}
S_{wall} \simeq  \int_{-\infty}^{+\infty}d x \,\left(V(\Phi) - {1\over 2}\Phi {d V \over d\Phi}\right)_{y=0}=
\int_{-\infty}^{+\infty}d x \,{\l\over 4}\left(v^4 -f_{kink}^4(x)\right).
\eeq

 In the 4D regime, in which eqs. (\ref{kinkeq2}) and  (\ref{kink2}) hold, then after an integration by parts we obtain:
\beq
S_{wall} = \pi R \int_{-\infty}^{+\infty} d x \,(\de_x f_{kink})^2 = {4 \over 3}\pi R  \,\mu \,v^2  = {4\over 3}\sqrt{\pi R \l } v^3
\eeq

Using this result in eq. (\ref{sbounce}) and minimizing the action
with respect to $\r_0$ we find:
\beq \r_0 \sim {S_{wall}\over b v}
\sim \sqrt{\pi R \l} { v^2 \over b} \eeq
 and the thin wall approximation holds if
\beq
1\ll \mu \rho_0 = {\l v^3\over b}
\eeq
which is the same condition \cite{coleman} finds in the purely 4D case,
and that we would have obtained had we started from the 4D effective
action in Section 2.1.1.

From the previous discussion, it is clear that the thin-wall approximation gets
worse and worse as we move away from the 4D regime, i.e. as $R \l v^2$ becomes large.
In fact, as discussed earlier, from  eqs. (\ref{lw}) and (\ref{l0}) it follow that
for  $R \l v^2>>1 $  the width of the wall becomes much larger than the size
of the region where the field profile has its largest variation (i.e. close
to the $f=0$), therefore the wall energy density gets spread over a larger
and larger region.

\section{The warped case} \label{warped}

We are now going to repeat the steps in the previous section in a
slice of $AdS_5$ bounded by two branes at $y=0,\pi R$. We
parametrize the metric as:
\beq\label{metric}
ds^2 = dy^2 + e^{-2 k
y} \eta_{\mu\nu} dx^\mu dx^\nu \ .
\eeq
We will always assume a
large warping, $\exp [k \pi R] \gg 1$.

The field equation is \beq \de_y^2 \Phi -4 k \de_y \Phi + e^{2ky}
\de_\mu^2 \Phi = 0 \ . \eeq with the same boundary conditions as
before. The general solution for a mass eigenstate, $\Box_4 \Phi =
m^2 \Phi$, has the form: \beq \Phi_m(y) = e^{2ky} \ B_2
\left({m\over k} e^{ky}\right) \eeq where $B_\nu(x) = a_m J_\nu(x) + b_m
N_\nu(x)$ is an appropriate  combination of Bessel functions, whose coefficients are
to be determined along with the mass eigenvalues $m$, from the
boundary conditions (\ref{bc0}-\ref{bc1}). The correct linear
combination is :
\beq\label{B2} B_\nu(x) = J_\nu(x) - {J_1(m e^{k\pi
R}/k) \over N_1 (m e^{k\pi R}/k)} N_\nu(x),
  \eeq
and the  resulting
equation for the KK masses is: \beq\label{kkwarped} m \
{B_1(m/k)\over B_2(m/k)} \ = \ \mu_0^2 \ . \eeq Typically the lowest
KK mass is of order:
\beq\label{mkk} m_{kk} \sim k e^{-k\pi R}.
\eeq
If we are in the 4D regime, when the lowest mode has  mass $m_0 <<
m_{kk}$, expanding the Bessels in (\ref{kkwarped}) we find: \beq
m^2_0 \sim {2 k \mu^2_0\over (1- e^{-2 k \pi R})} \eeq so the 4D
regime demands that \beq
 \l v^2 \ll k e^{-2 k \pi R} \ . \label{4dwarped}
\eeq

In this 4d regime, the wave function of the zero mode is basically
flat, whereas the $\Phi \rightarrow - \Phi$ symmetry breaking term
is redshifted by the warp factor $b \rightarrow b \exp (- 4 k \pi
R)$. The 4d Coleman expression for the bounce action is therefore
valid and produces a huge enhancement of the lifetime of the false
vacuum compared to the unwarped case. This is one of the main
advantages in constructing metastable vacua in warped spaces.

 However, it would be very interesting to also understand
the opposite regime,  namely  when $\mu_0^2  = \l v^2$ is much
larger than the KK scale. Notice that in this regime the mass
eigenstates are approximately given by the solutions of $B_2(m/k) =
0$, since in this case the l.h.s. of eq. (\ref{kkwarped}) is large.

Let us now look for a kink-like solution, depending on the coordinates
 $x$ and $y$.
From the flat case, we learned how to read-off an effective
one-dimensional equation for the field at $y=0$ from the spectral
equation. Repeating the argument that leads to (\ref{tan}), starting
from the general bulk solution with correct boundary condition at $y
= \pi R$
\bea
&& \Phi (x,y) \ = \ e^{2 k y} \ \int dp  \ a_p \ e^{p
x}  \left[ J_2 ({p \over k} e^{k y}) - {J_1({p \over k}  e^{k\pi R})
\over N_1 ({p \over k} e^{k\pi R})} N_2({p \over k} e^{k y}) \right]
\nonumber \\
&& = \ e^{2 k y} \ { {J_2 ({\de_x \over k} e^{k y}) - {J_1({\de_x
\over k} e^{k\pi R}) \over N_1 ({\de_x \over k} e^{k\pi R})}
N_2({\de_x \over k} e^{k y})} \over  {J_2 ({\de_x \over k}) -
{J_1({\de_x \over k} e^{k\pi R}) \over N_1 ({\de_x \over k} e^{k \pi
R})} N_2({\de_x \over k} )} } \ \Phi (x,0) \ , \label{warped1} \eea
 we obtain:
\beq \label{bessel} {B_1(m/k) \over B_2(m/k)}\Big|_{m\to \de_x}\de_x
f = \l f \left[f^2 -v^2\right] \ , \eeq where  $f(x) \equiv
\Phi(x,0)$.

In the 4D limit we can take the first term in the expansion of the l.h.s. of
eq. (\ref{bessel}), and we find again
a second order equation,  of the form:
\beq\label{kinkeq3}
 \de_x^2 f = {2k \l \over\left(1- e^{-2k\pi R}\right)} f \left[f^2 -v^2\right]
\eeq
which is the usual kink equation.

Let us estimate the width of the kink in the opposite regime,
$\l v^2 \gg m_{kk}^2/k $.
Using the same argument as in the previous section,
and writing $f(x) = \pm v + \eta e^{\pm x/l_w}$ we find that $l_w$ obeys:
\beq
l_w^{-1} {B_1(1/(k l_w))\over B_2(1/(k l_w))} = \mu_0^2,
\eeq
which comparing with eq. (\ref{kkwarped}) means that the maximal
 width is   equal to
the inverse mass of the lowest KK  mode, i.e. of order  (\ref{mkk}).

Let us now add the linear term (\ref{linear}) on the IR brane, as we did
in the flat case. Making the same approximations as in Section 2.2 (treat
$V_1$ as a perturbation, and use the thin-wall approximation), we
arrive at the following bounce action:

\bea\label{sbouncew}
{S_b\over 2\pi^2} &=& \int_0^{\pi R} dy e^{-4 k y} \int d\r \r^3 {1\over 2} \left[e^{2ky} (\de_\r \Phi^B)^2  + (\de_y \Phi^B)^2\right] \nn\\
&&  -   \int d\r \r^3 \left(V_0(\Phi^B(\r,0)) - e^{-4 k \pi R} b (\Phi^B(\r,\pi R)-v )\right)\nn \\
&& \approx -2  b v {\rho_0^4\over 4}e^{-4 k \pi R} + \rho_0^3 S_{wall},
\eea
where again  we have assumed that $\Phi^B(\r ,y) = \Phi^T(\r,y) = -v$ for
 $\r<\r_0$,   $\Phi^B(\r ,y) = \Phi^F(\r,y) = +v$ for  $\r>\r_0$,
and $\Phi^B(\r ,y) = \Phi_{kink}$ for $\r\approx r_0$. Notice the
appearance of the warp-factor in the first term of eq.
(\ref{sbouncew}) . $S_{wall}$ is the same as in eq. (\ref{swall}),
and it is localized on the brane at $y=0$ (we are neglecting the
subleading contribution from the IR brane to the wall energy).

 Minimizing eq. (\ref{sbouncew}) with respect to $\r_0$ we find:
\beq
\r_0 \sim {S_{wall}\over b v} e^{4 k \pi R}
\eeq
and if $S_{wall}$ is not too small this leads to an exponentially large
radius of the  vacuum bubble, and hence an exponentially small decay rate.

We can give a crude  estimate of $S_{wall}$ as follows.
Assume that $f_{kink}(x)$ can be approximated piece-wise
as:
\beq\label{approx1}
f_{kink}(x) \approx  \left\{ \begin{array}{ll} -v & x < -l_w, \\
                                        {v\over l_w} x \quad  & -l_w < x \simeq <l_w \\
                                         v & x >l_w  \end{array}\right.
\eeq
Then  evaluating the l.h.s. of  (\ref{swall}) with this approximation we find
\beq
S_{wall}\sim l_w \l v^4 \simeq {\l v^4\over k} e^{k\pi R}
\eeq
In practice this may be an overestimate, since the linear regime assumed
in (\ref{approx1}) may not be valid for the whole width of the wall. But
we can say that $S_{wall}$ is larger than just the contribution from
the linear region:
\beq \label{lowerb}
S_{wall} >  S_{min} \approx   l_{lin} \l v^4
\eeq
where $l_{lin}$ is the region around the origin where the linear
approximation (\ref{approx1}) is justified. It seems reasonable to believe
that this region is independent of $R$ for large enough $R$. In the
flat case this region is of the order $\mu_0^2 = \l v^2$ for large $\mu_0$.
We can repeat the same analysis of Section (\ref{5Deq}) to
estimate the slope of the solution  near  $x=0$. Assuming a behavior
of the type
$f(x) \sim \sin (x/l_0)$ we get the following equation:
\beq\label{l0warp}
 {i \over l_0}\  { {J_1 ({i \over k l_0 }) - {J_1({i  e^{k\pi R}
\over k l_0}) \over N_1 ({i  e^{k\pi R}\over k l_0})}
N_1({i  \over k l_0})} \over  {J_2 ({i \over k l_0}) -
{J_1({i e^{k\pi R}\over k l_0}) \over N_1 ({i  e^{k \pi
R}\over k l_0})} N_2({i\over kl_0} )} } = - \mu_0^2
\eeq
Now, let us assume that $k l_0 \gg 1$, so we can expand the Bessel functions
evaluated in $ (k l_0)^{-1}$ (but not the ones evaluated in
$ (k l_0)^{-1} e^{k\pi R}$ ).
 The quantity $J_1 (i  e^{k\pi R}/ k l_0 )/N_1 (i  e^{k\pi R}/ k l_0)$ is never
 small, since $J_1(ix)$ has no zeros outside the
origin. Using this fact, and expanding the Bessel
functions of argument $(kl_0)^{-1}$ {\em only}, we
obtain  to lowest order:
\beq\label{l0warp2}
{1\over k l_0^2} \simeq 2 \l v^2.
\eeq
This result was obtained under the  assumption $ k l_0 \gg 1$, therefore
it is valid if $\l v^2 \ll k$. This assumption is needed anyway,
since we are in curved space, with a curvature scale of order $k$,
and we are neglecting the backreaction of the scalar field $\Phi$ on the
background, as well as the contribution of $V(\Phi)$ to the
brane stress tensor.

Using $l_0$ from (\ref{l0warp2}) as an estimate of the width
of the linear region, we get a
 a more conservative lower bound on $S_{wall}$ from (\ref{lowerb}):
\beq \label{lowerb2} S_{wall} \simgeq  l_0 \l v^4  =\sqrt{\l\over k}
v^3   \eeq up to $O(1)$ coefficients. If this is the case the size
of the bounce is: \beq\label{rho0} \rho_0 \simgeq \sqrt{\l\over k} {
v^2 \over b} \exp[4 k \pi R] \eeq and the thin-wall approximation
holds if \beq \rho_0 / l_w \ \sim \ {v^2 \sqrt{k \l} \over b} \exp[3
k \pi R] \ \gg \ 1 \ , \eeq which is easily satisfied due to the
warp-factor. The decay rate is also exponentially suppressed:
plugging (\ref{rho0}) into (\ref{sbouncew}) we obtain: \beq S_b
\simeq {\l^2 v^9\over k^2 b^3} \exp[12 k \pi R] \eeq which gives a
huge lifetime $\tau= e^{S_b}$ even for moderate warping. This
estimate shows that,  in the warped case, we  don't need to restrict
to the 4-dimensional regime\footnote{As this regime demands  that
$\l v^2 \ll k \exp[-2k\pi R]$, this would impose a very strong
constraint on the model parameters. See however the next section for
a different model.} in order to have a small vacuum decay rate.

There is an important omission in our previous discussion, the
possible gravitational effects on the creation of the bubble.
Indeed, Coleman and de Luccia showed \cite{clucia} in the 4d context
that gravitational effects are negligible only in the case \beq {
\rho_0 \over \Lambda_0} \ \ll \ 1 \ , \label{warped5} \eeq where
$\Lambda_0$  is the radius such that the bubble radius equals the
Schwarzchild radius. For the 4d version of our model it equals
$\Lambda_0 = (16 G_N b v /3)^{-1/2}$, where $G_N = 1/M_P^2$ is the
4d Newton constant. In our case and when the 4d approximation is
valid, the two length scales scale with the warp factor as \beq
\rho_0 \rightarrow \exp[4 k \pi R] \ \rho_0 \quad , \quad \Lambda_0
\rightarrow \exp[2 k \pi R] \ \Lambda_0 \ . \label{warped6} \eeq
Then, neglecting factors of order one, gravity effects on the
creation of the bubble are negligible when \beq \exp[2 k \pi R] {1
\over M_P} \sqrt{\lambda v^5 \over b} \ll 1 \ . \label{warped7} \eeq
If the 4d limit (\ref{4dwarped}) is satisfied but (\ref{warped7}) is
violated, as shown in \cite{clucia}, there are two different cases.
In the first, the metastable vacuum has positive energy whereas the
true vacuum where we live has zero energy. Then gravity effects
increase substantially the probability of tunneling. In the second
case, the metastable vacuum has zero energy and tunnels into a
negative energy stable vacuum. In this case, gravity effects
increase the lifetime of the metastable vacuum. In the limit where
$\rho_0 > 2 \Lambda_0 $ the bubble cannot form and the metastable
vacuum becomes completely stable. This becomes therefore one
important outcome of having a warped extra dimension, in the case
where (\ref{warped6}) is violated.

\section{Supersymmetric extension : the AdS-ISS model} \label{susy}

Recently, there was a renewed interest in metastable vacua from the
point of view of supersymmetry breaking \cite{iss}, with further
applications to gauge mediation models \cite{gauge} and moduli
stabilization \cite{dpp}. The proposal in \cite{iss}
used the electro-magnetic Seiberg duality to argue for the existence
of metastable vacua in the supersymmetric QCD with a number of
flavors $ N_c+1< N_f < 3N_c/2$. In the IR free magnetic description
and before adding the effects of the (magnetic) gauge group, the
model is described by the O'Raifeartaigh-type model \beq W =
h q \Phi
{\tilde q} - h {\tilde \mu}^2 Tr \Phi \ , \label{s1} \eeq where
$q_i^a$ (${\tilde q}^{\bar j}_{\bar b}$) are the magnetic quarks
(antiquarks), $\Phi_{\bar j}^i$ are the mesons, $a,b= 1 \cdots N$
are color indices and $i,j= 1 \cdots N_f$ are flavor ones.
Supersymmetry is broken by the "rank condition", in the sense that
the supersymmetry condition \beq F_{\Phi} = h q {\tilde q} - h
{\tilde \mu}^2 I_{N_f} \ , \label{s2} \eeq where $I_{N_f}$ is the
$N_f \times N_f$ identity matrix, cannot be satisfied, since $q
{\tilde q}$ is a matrix of rank at most equal to $N<N_f$.
 One of the important requirements for the metastable vacuum to be
long-lived in the ISS model is $\epsilon \equiv {\tilde \mu} /
\Lambda_m << 1$, where $\Lambda_m$ is the Landau pole of the
magnetic theory. From a string theory viewpoint \cite{issstrings},
one natural realization of the ISS model,in its magnetic
description, is in terms of D3/D7 brane configurations, with the ISS
gauge group realized on the D3 branes, with (anti) quarks coming
from the D3-D7 sector and the magnetic mesons being the positions of
a stack of D7 branes.

The purpose of this section is to analyze in a field-theoretical
example the effect that the warping of the internal space, generated
by the branes, could have on the model. We model this effect by
considering a five-dimensional supersymmetric model in a slice of
$AdS_5$ \cite{rs} with the metric \beq d s_5^2 = e^{- 2 k  |y|}
\eta_{\mu \nu} dx^{\mu} dx^{\nu} + d y^2 \ , \eeq  with ISS gauge
fields and the quarks, antiquarks confined to the UV boundary $y =
0$ and the mesons promoted to a hypermultiplet $(\Phi_1,\Phi_2)$
propagating into the 5d bulk, with $Z_2$ parities $(+,-)$. The
mesons-quark coupling is localized on the UV brane, whereas we
choose to put the linear term in the ($Z_2$ even) mesons $\Phi_1$ in
the superpotential on the IR brane\footnote{A  geometrical
construction in a string context, similar in spirit,  was proposed
in \cite{absv}.}. As we will show below,  due to the exponential
warp factor, the mass parameter ${\tilde
\mu}$  will be redshifted
such that the lifetime of the metastable vacuum becomes
arbitrarily large. In a manifest 4d supersymmetric language
\cite{agw,gp}, the Lagrangean describing the system is \bea &&S =
\int d^4 x dy \left\{ \int d^4 \theta \ e^{- 2 k y}
(\Phi_1^{\dagger} \Phi_1 + \Phi_2^{\dagger} \Phi_2 ) + \int d^2
\theta e^{- 3 k y}
(\Phi_2 \partial_y \Phi_1 + {\rm h.c})\right. \nonumber \\
&& + \left[ \int d^4 \theta \ (q^{\dagger }q + {\tilde q}^{\dagger}
{\tilde q} ) + \int d^2 \theta \ \left(h q \Phi_1 {\tilde q} + W_{np}
(\Phi_1) + {\rm h.c}\right) \right] \ \delta
(y) \nonumber \\
&& \left.- \left[\int d^2 \theta \ e^{- 3 k \pi R} (\ h {\tilde \mu}^2 \Phi_1 +
{\rm h.c})\right] \ \delta (y - \pi R) \ \right\} \ , \label{s3} \eea where
$W_{np}$ is the non-perturbative mesonic superpotential arising in
the field direction where the mesons $\Phi_1$ get vev's, give masses
to the quarks (antiquarks) and generate the IR dynamics restoring
supersymmetry.
 The (metastable) supersymmetry breaking becomes now a non-local effect and arises do
to the impossibility, in the absence of $W_{np}$, to solve the
supersymmetric condition:
\beq e^{- 2 k y} F_{\Phi_1} \ = -\de_y \left( e^{- 3 k
y}  \Phi_2\right) + \left(h q {\tilde q} + \de_{\Phi_1} W_{np}\right) \ \delta (y)-  e^{- 3 k \pi R
} \ h {\tilde \mu}^2 I_{N_f} \delta (y-\pi R) \ . \label{s4}
\eeq
In order to cancel the last term in eq. (\ref{s4}), the $Z_2$-odd
mesons $\Phi_2$ acquires a non-trivial profile:
\beq\label{s4b}
\Phi_2 = e^{3k(y-\pi R)}(h {\tilde \mu}^2 /2) I_{N_f} \epsilon (y).
\eeq
 If $W_{np}=0$  supersymmetry is  broken:
with (\ref{s4b}), and using \mbox{$\de_y \epsilon(y) = 2 [\delta(y)-
\delta(y-\pi R)]$}, eq. (\ref{s4}) becomes:
\beq\label{s4c}
 e^{- 2 k y} F_{\Phi_1} \ = \delta(y)\left[h q {\tilde q} -  e^{- 3 k \pi R
} \ h {\tilde \mu}^2 I_{N_f}\right], \eeq which cannot vanish due to
the rank condition. Notice that the parameter which controls
supersymmetry breaking is not $\tilde{\mu}$, but rather
\beq\label{mueff} \mu_{eff}^2 = e^{-3k \pi R} \tilde{\mu}^2 \ , \
{\rm since} \quad q {\tilde q} = e^{-3k \pi R} \tilde{\mu}^2 I_N \ .
\eeq

 The presence of $W_{np}$ restores supersymmetry by producing
sources which do add up to zero. From the point of view of the bulk
fields, in the metastable vacuum $\Phi_1$ gets a boundary mass term
\beq \mu_0^2 = h^2  \langle q^{\dagger }q + {\tilde q}^{\dagger}
{\tilde q} \rangle = 2 h^2 N  \ e^{-3k \pi R} \tilde{\mu}^2 \ ,
\label{s5}\eeq whereas in the supersymmetric vacuum it gets also
localized nonperturbative interactions. The formally divergent terms
$\delta (0)$ in (\ref{s3}) do not appear in physical quantities, as
shown in various similar situations \cite{mp}.

Notice the close analogy of this model with the toy model analyzed
in section 3: the symmetry breaking parameter is redshifted by a power
of the scale factor. However  in this model the validity of the
4D limit is automatic,
and does nor require an additional fine tuning: the
existence of a light mode for $\Phi_1$ requires
$\mu^2_0 << k \exp[-2kr]$ , and from eq. (\ref{s5}) we see that
this does not impose any strong constraint on $h$ and $\tilde{\mu}$,
provided the warp factor is large. Therefore, since the 4D
limit analysis holds, the smallness of the symmetry-breaking parameter
due to the redshift  leads immediately
to an exponential enhancement of the lifetime of the
metastable vacuum.
 There is one  critical point to check: this conclusion is valid
if  the wave function of
the lightest mode of $\Phi_1$ does  not grow too fast in the IR and
destroys the redshift of the mass term ${\tilde \mu}$, transparent
in (\ref{s3}). In the limit where the 4d effective theory is valid,
i.e. the lightest mode is much lighter than the KK masses
$m << k e^{- k \pi R}$, its corresponding wave function reads
approximatively \beq \Phi_1^{(0)} (y) \ \simeq \ d_1 \ e^{4 k y} \left[ 1
- {m^2 \over 12 k^2} e^{2 k y} \right] + d_2 \ \left[ 1 + {m^2 \over 4 k^2} e^{2
k y} \right] \ . \label{s6} \eeq Boundary conditions determine then the
mass spectrum to be given by the equation \beq - 2 m^2 e^{- 2 k \pi
R} = \left(\mu_0^2 - {m^2 \over 2k}\right) \left(4k - {m^2 \over 2k} e^{2 k \pi R}\right)
\ . \eeq Due to the validity of the 4D limit,
we get the 4D result (see section 3) $m^2 \simeq 2 k
\mu_0^2$ and a corresponding wavefunction (\ref{s6}) which is
constant in $y$ to the leading order. In this case, the
redshift  of the mass
parameter ${\tilde \mu}^2 \rightarrow {\tilde \mu}^2 e^{- 3 k \pi
R}$ is effective and produces a huge enhancement of the lifetime of
the false vacuum. Notice that with respect to a 5d flat metric, the
light mode is actually localized on the UV boundary. Since the KK
modes and the linear term are localized on the IR boundary, this
explains the enhancement of the lifetime of the metastable
vacuum.

Another interesting case, with the same matter content, is when the
whole superpotential is localized on the UV boundary. In this case,
there is no redshift of the mass parameter ${\tilde \mu}$ and
generically no light mode. One way to obtain a light mode even in
the case $ k e^{- 2 k \pi R} << \mu_0^2 << k$ is to add a bulk mass
for the hypermultiplet, which is tuned appropriately against the
boundary mass. This is a tuning in a non-supersymmetric setup, but
the tuning is actually required and protected versus radiative
corrections by supersymmetry \cite{gp,tony}. In this case the bulk
mass $m_b$ and the boundary masses $\mu_0$, $\mu_{\pi}$ for the
scalar component of $\Phi_1$, in the false vacuum, are given by \bea
&& {m_b^2\over k^2} \ = \ \alpha^2 - 4 \ = \ \left(c-{3 \over
2}\right) \left(c+{5 \over 2}\right)
\ , \nonumber \\
&& \mu_0^2 \ = \ h^2  \langle q^{\dagger }q + {\tilde q}^{\dagger}
{\tilde q} \rangle + \left({3 \over 2} -c\right) \ k \ , \nonumber \\
&& \mu_{\pi}^2 \ = \  - \ \left({3 \over 2} -c\right) \ k \ ,
\label{s7} \eea where $\alpha = |c+1/2|$. Since we want to preserve
in the first approximation the $AdS_5$ geometry, we are interested
in small backreaction of the scalar field and therefore small bulk
mass $\alpha \simeq 2$. There is one interesting example of this
type, with $c = - 5/2 $ and therefore zero bulk mass for $\Phi_1$,
with non-vanishing brane localized masses. In this case we find a
light scalar mode localized on the IR brane, with wave-function and
mass given by \bea
&& \Phi_1^{(0)} (x,y) \ \sim  \ e ^{- 3 k \pi R}   e^{4  k y} \phi (x) \ , \nonumber \\
&& m^2 \simeq 6 k h^2  \ \langle q^{\dagger }q + {\tilde
q}^{\dagger} {\tilde q} \rangle \ e^{- 6 k \pi R} \ . \label{s8}
\eea The term $\exp(- 3 k \pi R)$, important in what follows, comes
from normalization of the 4d kinetic term of the light mode
$\phi(x)$. The four dimensional Lagrangean in this case is very
close to the 4d ISS Lagrangean. Auxiliary fields are \bea && e^{- 2
k y} F_{\Phi_1} =  - \de_y (e^{- 3 k y} \Phi_2) +
(h q {\tilde q} - h {\tilde \mu}^2 + \de_{\Phi_1} W_{np}) \delta (y) \ , \nonumber \\
&& F_q = e ^{- 3 k \pi R} \phi {\tilde q} \quad , \quad F_{\tilde q}
= e ^{- 3 k \pi R} q \phi \ .
 \eea
Therefore, due to the wave-function in (\ref{s8}), the meson-quark
coupling gets changed and become \beq e ^{- 6 k \pi R} \ h^2
|\phi|^2 \ \left( |q|^2 + |{\tilde q}|^2 \right) \ . \label{s08}
\eeq
In the ISS vacuum, the quark vev's are as in 4d
\begin{equation}
q  \ = \ {\tilde q}^T \ =
 \left(
\begin{array}{c}
\mu I_N \\
0
\end{array}
\right)
 \ , \label{iss3}
\end{equation}
Then (\ref{s08}) reproduces the light meson
mode (\ref{s8}). In the SUSY vacuum in which mesons get vev's, quark
masses are also redshifted by the same factor $m_q^2 = m_{\tilde
q}^2 = \exp(- 6 k \pi R) h^2 |\phi|^2 $. Therefore the distance in
field space between the ISS and the SUSY vacuum is greatly enhanced
$\Delta \phi = \exp [3 k \pi R N/(N_f-N)] \Delta \phi_{ISS}$,
whereas the barrier remains unchanged $V_{peak} = N_f h^2 {\tilde
\mu}^4$. Therefore the bounce action $S_b $ in the triangular
approximation $S_b \sim (\Delta \phi)^4 / V_{peak}$ \cite{dj} and
the lifetime of the false vacuum are accordingly increased \beq S_b
\ \rightarrow \ e ^{ 12 k \pi R N \over N_f-N} S_b \ .  \eeq

However, as discussed in the previous section, a more detailed
analysis of gravitational effects is needed in order to check if they
are negligible. Again, if the metastable vacuum has zero energy
whereas the stable vacuum has negative one, one expects the lifetime
to be increased and eventually the false vacuum to become completely
stable \cite{clucia}.

Notice that for values $h^2 {\tilde \mu}^2 \sim k $ and by defining
the mass scale on the IR brane with a dynamical scale $\Lambda
\equiv k \exp (- k \pi R)$, we can rewrite qualitatively (\ref{s8})
in the suggestive way \beq m \sim { \Lambda^3 \over M_P^2} \ ,
\label{s9} \eeq where $M_P$ is the 4d Planck mass. It is interesting
to notice the analogy between (\ref{s9}) and the scale of
supersymmetry breaking in the observable sector in ${\cal N}=1$
supergravity with a gaugino condensation $\langle \l \l \rangle =
\Lambda^3$ in a hidden sector, coupled gravitationally with the
observable one.

The models presented here  can  be interpreted from a holographic
point of view. The metastable susy breaking can be understood in a
purely four-dimensional way as arising from the infrared dynamics of
a strongly coupled CFT sector, dual to the bulk geometry and the
bulk fields $\Phi_{1,2}$. This CFT acts as a hidden sector,  coupled
to the quarks living on the UV brane. In the first example presented
in this section (zero bulk mass), the light mode mediating the
vacuum decay is localized  on the UV brane, and from the point of
view of the 4D theory it is a fundamental degree of freedom. The
redshift of the mass parameter ${\tilde \mu}$ could be interpreted
as the holographic version of the retrofitting discussed in
\cite{dfs,brummer}.
 In the second example, in which  the bulk field profile is given by eq.
(\ref{s8}),   the light mode is peaked  on the IR brane and couples only
gravitationally to the UV brane.
In both cases, in the holographic 4D theory description the symmetry
breaking occurs as an infrared effect, generating a hierarchy of scales
like in eqs. (\ref{mueff}) and (\ref{s9})


In other types of models, in the nontrivial limit in which the
boundary masses are large and the KK modes are expected to play a
role in the bounce, there is no light mode anymore in the spectrum
and the methods of Section 3 are needed in order to estimate the
lifetime of the false vacuum.
 Finally, we would like to point out that there is nothing peculiar
about the ISS model from the point of view of a phenomenological
construction in a 5d warped space. Traditional O'Rafeartaigh models
can be similarly discussed, with corresponding mass parameters and
consequently scale of supersymmetry breaking  redshifted to very
small values. Since our main motivation was to understand the
properties of the classical kink and bounce solutions, we refrain
ourselves to discuss further here these applications.


\section*{Acknowledgments}{ We would like to thank A. Hebecker,
E. Kiritsis and especially G. Pradisi for useful discussions. E.D
thanks the Univ. of Warsaw
for hospitality during the completion of this work. Work partially
supported by the CNRS PICS \#~2530 and 3059, RTN contracts
MRTN-CT-2004-005104 and MRTN-CT-2004-503369, the European Union
Excellence Grant, MEXT-CT-2003-509661 and the EC contract
MTKD-CT-2005-029466. F.N. is supported by an European Commission Marie Curie
Intra European Fellowship, contract
 MEIF-CT-2006-039369.}

\appendix
\renewcommand{\theequation}{\thesection.\arabic{equation}}
\addcontentsline{toc}{section}{Appendices}

\section{Standard Kink solution} \label{appkink}

Here, we remind the reader of the standard domain wall, or kink,  solution.
Consider  a scalar field with quartic potential,
\beq
 V(\Phi) = {\l\over 4} \left(\Phi^2 -v^2\right)^2 \ . \label{k1}
\eeq
The one-dimensional field equation,
\beq\label{intro1}
  \Phi''(x) = {dV\over d\Phi} \ = \ \l \Phi (\Phi^2 - v^2) \ ,
\eeq with boundary conditions $\Phi(-\infty)=-v$, $\Phi(+\infty)=v$
is solved by: \beq\label{intro2} \Phi_{kink}(x) = v \tanh \mu x,
\qquad \mu\equiv \sqrt{\l v^2\over 2}. \eeq This is also a solution
of  the  first order equation:
\beq\label{intro3}
{\cal E}\equiv {\left( \Phi'
\right)^2 \over 2} \ - \ V(\Phi)=0 \ .
\eeq
This can be read as the conservation of energy equation   of
a point particle moving in the potential $-V$ with vanishing total ``energy''
 ${\cal E}$.

The total energy of the
kink is
\beq
E \ = 2\ \int_{-\infty}^{\infty} dx V (\Phi) \ = \
{2\sqrt{2 \l} \over 3} \ v^3 ={16\over 3}V(0){1\over \mu} \ .
\eeq


\section{Conserved energy}

In this appendix we derive the conserved "energy," eq. (\ref{ene}).
We start from equation (\ref{tan}) which we write in the form
\beq
\sum_{n=1}^{\infty}a_{n}\partial ^{2n} f-V'(f)=0,\label{exp}
\eeq
where the $a_n$ are defined by $\tan(\pi R x)=\sum_n a_n x^{2n-1}$.
 Next we multiply (\ref{exp}) by $\partial f$ and  use the following identity
 \beq
 \partial f \partial ^{2n} f={1\over 2}
 \partial \sum_{p=1}^{2n-1}(-1)^{p+1}\partial^p
 f\partial ^{2n-p}f
\eeq
to get
\beq
\partial[{1\over 2}\sum_{n=1}^{\infty}a_{n}
  \sum_{p=1}^{2n-1}(-1)^{p+1}\partial^p
 f\partial ^{2n-p}f- V(f)]=0.
 \eeq
 We deduce the conserved quantity
 \beq
 \pi R{\cal E}={1\over 2}\sum_{n=1}^{\infty}a_{n}
  \sum_{p=1}^{2n-1}(-1)^{p+1}\partial^p
 f\partial ^{2n-p}f- V(f).\label{inter}
 \eeq
 Symbolically the sum $\sum_{p=1}^{2n-1}(-1)^{p+1}\partial^p
 f\partial ^{2n-p}f$ can be written as
 \beq
 {\partial_1\partial_2\over \partial_1+\partial_2}(\partial _1^{2n-1}+
 \partial_2^{2n-1}) f(x_1)f(x_2)|_{x_1=x_2=x},
 \eeq
 where
 $\partial_{i}=\partial_{x_i}$. The first term
 in (\ref{inter})  can thus be put in the form
 \beq
 {\partial_1\partial_2\over \partial_1+\partial_2}(\tan(\pi R\partial_1)+
 \tan(\pi R\partial_2)) f(x_1)f(x_2)|_{x_1=x_2=x}
\eeq
Now we use
\beq
\tan(\pi R\partial_1)+
 \tan(\pi R\partial_2)=[1-\tan(\pi R\partial_1)\tan(\pi R\partial_2)]
 \tan(\pi R(\partial_1+\partial_2))
 \eeq
 and $(\partial_1+\partial_2)^nf(x_1)f(x_2)|_{x_1=x_2=x}=\partial^n f^2$,
 which gives
 \beq
 \tan(\pi R(\partial_1+\partial_2))f(x_1)f(x_2)|_{x_1=x_2=x}
 =\tan(\pi R\partial)f^2.
 \eeq
 Collecting all the terms we get the final expression
 \beq \label{conserved}
 \pi R{\cal E}={1\over 2}\left[{\tan(\pi R\partial)\over \partial}\right]
 \left[(\partial f)^2-(\partial\tan(\pi R\partial) f)^2\right]-V(f).
 \eeq
\section{The 5D Kink near the extrema of the scalar potential} \label{fixed}

Here we analyze the behavior of the kink  in flat space,
close to the extrema of the potential, $f=0,\pm v$. In particular
we show that a solution with zero ``energy'' ${\cal E}$, i.e. satisfying
eq. (\ref{ener}), cannot cross from a region where $|f|<v$ to another
one where $|f|>v$.

We have already shown in Section \ref{5Deq} that when $f \sim v$ the solution
to eq. (\ref{tan}) is exponential,
\beq \label{lwapp}
 f \sim v \pm  \eta \exp[\pm x/l_w], \qquad   {1\over l_w} \ \tan \left({\pi
R\over l_w}\right) \ = \ 2 \l v^2 \ ,
\eeq
where $\eta$ is a constant.
One can check that the above ansatz  satisfies the condition ${\cal E}=0$
to lowest order  in $\eta$: inserting (\ref{lwapp}) in
(\ref{ene}) and keeping terms quadratic in $\eta$ we obtain (for any choice
of signs in (\ref{lwapp})):
\bea
{\cal E} && = {1\over 2}{\tan(\pi R\partial) \over \pi R \partial}
\left[\left({\pm \eta\over l_w}e^{\pm x/l_w} \right)^2-\left(\tan(\pi R/l_w){\pm \eta\over l_w}e^{\pm x/l_w} \right)^2\right]-{\l v^2 \eta^2  \over \pi
R}e^{\pm 2x/l_w} \nn\\
&& = \Bigg\{{1\over 2}{\tan(2 \pi R/l_w ) \over 2 \pi R/l_w}
\left[{1\over l_w^2}\left(1- \tan^2(\pi R/l_w)\right) \right]-{\l v^2  \over \pi
R}\Bigg\}\eta^2  e^{\pm 2x/l_w}  \nn\\
&& = \Bigg\{{1\over l_w} \tan\left({\pi R\over l_w}\right)- 2\l v^2\Bigg\}{\eta^2  e^{\pm 2x/l_w}\over  2\pi R}\ = 0  , \nn
\eea
where in the last line  we used the identity:
\beq
\tan 2z = {2 \tan z\over 1-\tan^2 z}.
\eeq
Each  solution of the type  (\ref{lwapp}) approaches $\pm v$ as  $|x|\to \infty$. One can ask whether it is possible for a solution to approach (and cross)
$|f|=v $ at a finite value $x=x_0$. A priori, one can take :
\beq\label{super}
f \sim v + \eta \sinh [(x-x_0)/l_w] \qquad x\approx x_0\ ,
\eeq
as a solution to the
linearized kink equation (\ref{tan}) with the desired property
to cross  $f=v$ at $x=x_0$. However, let us compute the conserved
energy for (\ref{super}):
\bea
{\cal E} && = {1\over 2}{\tan(\pi R\partial) \over \pi R \partial}
\left[\left({1\over l_w} \cosh [(x-x_0)/l_w] \right)^2-\left(\tan(\pi R \de)
\de \sinh[(x-x_0)/l_w] \right)^2\right] \eta^2 \nn\\
&& - \l v^2 \eta^2 \sinh^2[(x-x_0)/l_w].
\eea
We will  use the following formal identity: for any differential operator
 $\hat{O}(\de)$ constructed with a function $O(k)$
 which has an expansion containing only  even powers of $k$,
(such as the two operators appearing in the above expression) we
have:
\beq
\hat{O}(\de) \sinh k x = O(k)\sinh k x, \quad  \hat{O}(\de) \cosh k x = O(k)\cosh k x \ .
\eeq
Using this fact, and some manipulation of the hyperbolic functions,
we arrive at :
\beq \label{nonzero}
{\cal E}  =  \left\{
{1\over 4 l_w^2}\left(1 + \tan^2 \left({\pi R\over l_w}\right)\right)
+ {\l v^2 \over 2\pi R}  \right\} \eta^2  > 0 \ ,
\eeq
therefore a solution that crosses $f=\pm v$ cannot have a zero
value of ${\cal E}$.

On the contrary, a zero energy solution can cross $f=0$ at some finite
value of $x$. Close to $f=0$ the solution of the
linearized equation has now the form (see eq. (\ref{l0})):
\beq\label{cross}
f(x) \simeq \eta \sin[(x-x_0)/l_0], \qquad \tanh[\pi R/l_0] = \l v^2 l_0,
\eeq
Inserting this in  eq. (\ref{conserved}) and performing the same steps that
led to eq. (\ref{nonzero})  we
obtain:
\beq
{\cal E} =     \left\{
{1\over 4 l_0^2}\left(1 - \tanh^2 \left({\pi R\over l_0}\right)\right)
+ {\l v^2 \over 2\pi R}  \right\} \eta^2 - {\l v^4\over 4\pi R}.
\eeq
The last term is $\eta$-independent, and comes from the non-zero value
of $V(f)$ at $f=0$. For an appropriate choice of $\eta$, we can make ${\cal E}$
vanish:
\beq
\eta^2 = {\l v^4\over 4\pi R} \left\{
{1\over 4 l_0^2}\left(1 - \tanh^2 \left({\pi R\over l_0}\right)\right)
+ {\l v^2 \over 2\pi R}  \right\}^{-1} \quad \Rightarrow \quad {\cal E} = 0
\eeq
Notice that this argument  does not require $\eta$ to be small: the validity
of the linearized approximation  made in eq.  (\ref{cross}) holds  for
arbitrary $\eta$, as long as $x$ is close enough to $x_0$.

\end{document}